\documentclass[pra,twocolumn,tightenlines,showpacs,nofootinbib]{revtex4}
\usepackage{multirow}
\usepackage{bm,dcolumn,amsmath,graphicx}

\newcommand{\eref}[1]{Eq.~(\ref{#1})}
\newcommand{\tref}[1]{Table~\ref{#1}}

\begin{document}
\title{Relativistic calculations of $C_6$ and $C_8$ coefficients for strontium dimers}
\author{S.~G.~Porsev$^{1,2}$}
\author{M.~S.~Safronova$^{1,3}$}
\author{Charles W. Clark$^3$}
\affiliation{ $^1$Department of Physics and Astronomy, University of Delaware, Newark, Delaware 19716, USA\\
 $^2$Petersburg Nuclear Physics Institute, Gatchina, Leningrad District, 188300, Russia \\
 $^3$Joint Quantum Institute, National Institute of Standards and Technology and \\
 the University of Maryland, Gaithersburg, Maryland, 20899, USA}
\date{\today}

\begin{abstract}
 The electric dipole and quadrupole polarizabilities of the $5s5p~^3\!P_1^o$ state
and the $C_6$ and $C_8$ coefficients for the $^1\!S_0 +\, ^1\!S_0$ and $^1\!S_0 +\, ^3\!P_1^o$ dimers of strontium are calculated using a high-precision relativistic approach that combines configuration interaction and linearized coupled-cluster methods.
Our recommended values of the long range dispersion coefficients for the $0_u$
and $1_u$ energy levels are $C_6(0_u)=3771(32)$~a.u. and
$C_6(1_u)= 4001(33)$ a.u., respectively.  They are in good agreement with recent results from experimental photoassociation data.
We also calculate $C_8$ coefficients for Sr dimers, which are needed for precise determination of long-range interaction potential.
We confirm the experimental value for the magic wavelength, where the Stark shift on the $^1\!S_0$-$^3\!P_1^o$ transition
vanishes. The accuracy of calculations is analyzed and uncertainties are assigned to all quantities reported in this work.
\end{abstract}
\pacs{34.20.Cf, 32.10.Dk, 31.15.ac}

\maketitle
\section{Introduction}
The divalent alkaline-earth element strontium
 is of interest for many applications of atomic, molecular and optical physics.
The atomic clock based on the $5s^2\,^1\!S_0 - 5s5p\,^3\!P_0^o$ transition in Sr has achieved a total systematic uncertainty
of $6 \times 10^{-18}$~\cite{BloNicWil14}, which is the smallest yet demonstrated.
The architecture of this clock also provides capabilities for detailed studies  of quantum many-body physics.
The SU(N)-symmetric interactions of $^{87}$Sr atoms in optical lattices provide a platform
for quantum simulation of lattice gauge theories and a variety of quantum materials such as  transition
metal oxides, heavy fermion compounds, and exotic topological phases.
Early demonstrations of this capability have been realized by high-resolution
spectroscopy of SU(N)-symmetric interactions in Sr orbital magnetism \cite{ZhaBisBro14}.

All four stable isotopes of Sr have been brought to strong quantum degeneracy:
Bose-Einstein condensation have been achieved  in the bosonic isotopes 84, 86 and 88, and
fermionic $^{87}$Sr has been cooled to within 10\% of its Fermi temperature \cite{PhysRevA.87.013611}.  The first isotope to be
condensed \cite{PhysRevLett.103.200401}, $^{84}$Sr, is also distinctive in being the only atomic species to date which has been
condensed by laser cooling alone \cite{PhysRevLett.110.263003}. Sr has also been used  in ultracold  gases of both
homonuclear \cite{SkoMozKoc12,ReiOsbMcD12} and heteronuclear \cite{TomPawKoc11} molecules. A quantum degenerate gas mixture of Sr and Rb has been
realized recently~\cite{PasBayTza13}, as a prerequisite for the production of a quantum degenerate gas of polar molecules.
Presently, there is much interest in Sr  photoassociation spectroscopy due to its relevance for the production of ground
state ultracold molecules \cite{SkoMosKoc14}, coherent photoassociation \cite{YanDesHua13}
and search for time-variation of the  electron-proton mass ratio \cite{KotZelYe09}.

Understanding of long-range interaction of the Sr atoms is needed for all of the applications mentioned above. In Ref.~\cite{ZhaBisBro14}, we have provided recommended values of the $C_6$ long-range interaction coefficients for the $^1\!S_0-\,^1\!S_0$, $^1\!S_0-\,^3\!P_0^o$, and
$^3\!P_0^o-\,^3\!P_0^o$ dimers for the determination of relevant interaction parameters for spin-orbital quantum dynamics.

Motivated by the diverse applications and particular interest of $^1\!S_0 - \,^3\!P_1^o$ intercombination line
for most recent photoassociation studies, we have calculated the $C_6$ and $C_8$ van der Waals coefficients for the Sr $^1\!S_0 +\, ^1\!S_0$ and $^1\!S_0 +\, ^3\!P_1^o$ dimers. The ground state long-range interaction coefficients were
previously studied in Refs.~\cite{PorDer02,MitBro03,PorDer06Z}; here we provide revised values that have been critically evaluated for accuracy.
In a recent paper, Borkowski {\it et al.}~\cite{BorMorCiu14} reported photoassociation spectroscopy of ultracold Sr atoms near
the intercombination $^1\!S_0 +\, ^3\!P_1^o$ line. They obtained the Coriolis mixing angles and linear Zeeman coefficients
for all of the photoassociation lines and determined the van der Waals $C_6$ coefficients for the $0_u$
and $1_u$ bound state energies to be $C_6(0_u) = 3868(50)$~a.u. and $C_6(1_u) = 4085(50)$ a.u..
Our recommended values of $C_6(0_u)=3771(32)$ a.u. and $C_6(1_u)= 4001(33)$~a.u.
provide further confidence in the fitting of precision photoassociation data.

In the course of our work, we also calculated a number of $E1$ transition amplitudes and  the electric dipole and quadrupole polarizabilities of the $5s5p~^3\!P_1^o$ state of atomic Sr for use in other applications. We report recommended values of these quantities here.

This paper is organized as follows. In Sec.~\ref{Method} we briefly describe the method of calculation and
present the matrix elements of $E1$ transitions from the $^3\!P_1^o$ state to low-lying even-parity states.
In Sec.~\ref{Polariz1} we discuss calculation of the scalar static $^3\!P_1^o$ polarizability. The $^1\!S_0-\,^3\!P_1^o$ magic
wavelength is discussed in Sec.~\ref{mw}.
Section~\ref{Sec:C6} is devoted to calculation of the van der Waals $C_6(^1\!S_0 +\, ^3\!P_1^o)$
coefficients. In Sections \ref{C8} and \ref{C8sub} we present the results of calculation of the electric quadrupole $^3\!P_1^o$ polarizability
and $C_8(^1\!S_0 +\, ^3\!P_1^o)$ coefficients, respectively.

\section{Method of calculation and electric-dipole matrix elements}
\label{Method}
We consider atomic Sr as an atom with frozen Ag-like Sr$^{2+}$ core and two valence electrons. Interaction of the valence electrons
is taken into account in the framework of configuration interaction (CI) method (see, e.g.,~\cite{KotTup87}) while
core-core and core-valence correlations are treated in the framework of many-body perturbation theory (MBPT)
and all-order single-double coupled-cluster method. Both CI+MBPT and
CI+all-order methods were described in detail in a number of papers~\cite{DzuFlaKoz96b,Koz04,SafKozJoh09,SafKozCla11}, so here
we only briefly review their main features. While the CI+all-order method is more accurate, carrying out the calculations
by both approaches allows us to estimate the accuracy of the final results.

Unless stated otherwise, we use atomic units (a.u.) for all matrix elements and polarizabilities throughout this
paper: the numerical values of the elementary charge, $|e|$, the reduced Planck constant, $\hbar = h/2 \pi$,
and the electron mass, $m_e$, are set equal to 1. The atomic unit for polarizability can be
converted to SI units via $\alpha/h$~[Hz/(V/m)$^2$]=2.48832$\times10^{-8}\alpha$~(a.u.), where the
conversion coefficient is $4\pi \epsilon_0 a^3_0/h$ and the Planck constant $h$ is factored out in
order to provide direct conversion into frequency units; $a_0$ is the Bohr radius and $\epsilon_0$
is the electric constant.

We start with the  solutions of  the Dirac-Fock equation
\begin{equation}
H_0\, \psi_c = \varepsilon_c \,\psi_c,
\label{H0}
\end{equation}
where $H_0$ is the Dirac-Fock Hamiltonian
and $\psi_c$ and $\varepsilon_c$ are single-electron wave functions and energies. The calculations are
carried out in the $V^{\rm{N-2}}$ potential, where $N$ is the total number of electrons and an initial self-consistent Hartree-Fock
procedure is applied to the $N-2=36$ core electrons. The wave functions and the energy levels for the valence electrons are determined
by solving the multiparticle relativistic equation ~\cite{DzuFlaKoz96b},
\begin{equation}
H_{\rm eff}(E_n) \Phi_n = E_n \Phi_n
\label{Heff}
\end{equation}
with the effective Hamiltonian defined as $$ H_{\rm eff}(E) = H_{\rm FC} + \Sigma(E).$$
Here $H_{\rm FC}$ is the Hamiltonian in the frozen-core approximation and the operator $\Sigma(E)$, accounting for virtual core
excitations, is constructed using second-order perturbation theory in the CI+MBPT method~\cite{DzuFlaKoz96b} and
using a linearized coupled-cluster single-double method in the CI+all-order approach~\cite{SafKozJoh09}.
Since the valence space contains only two electrons, the CI can be made numerically complete.
Our calculation of the energy levels was presented and discussed in detail in Ref.~\cite{SafPorSaf13}.
In analogy with the effective Hamiltonian we can construct effective electric-dipole and electric-quadrupole
operators to account for dominant
core-valence correlations~\cite{DzuKozPor98,PorRakKoz99P,PorRakKoz99J}.
\begin{table}
\caption{The CI+MBPT and CI+all-order results for absolute values of the reduced electric dipole matrix elements for the transitions that
give dominant contributions to the polarizabilities of the $5s5p\,^3\!P_0^o$ and $5s5p~^3\!P_1^o$ states. The CI+MBPT and
CI+all-order results including RPA corrections are given in columns labeled ``CI+MBPT'' and ``CI+All'', respectively.
The relative differences between the CI+all-order and CI+MBPT results are given in column labeled ``HO'' in \%.
The recommended values of the matrix elements are given in the last column (see text for details).}
\label{tab1}
\begin{ruledtabular}
\begin{tabular}{lcccc}
\multicolumn{1}{c}{Transition}    & CI+MBPT  & CI+All  &   HO     &  Recomm.\\
\hline \\[-0.5pc]
$5s5p~^3\!P_0^o - 5s4d~ ^3\!D_1$  &  2.681   &  2.712  &  1.14\%  &  2.675(13)$^{\rm a}$\\
$5s5p~^3\!P_1^o - 5s4d~ ^3\!D_1$  &  2.326   &  2.354  &  1.19\%  &  2.322(11)$^{\rm a}$\\
$5s5p~^3\!P_1^o - 5s4d~ ^3\!D_2$  &  4.031   &  4.075  &  1.08\%  &  4.019(20)\\[0.3pc]

$5s5p~^3\!P_0^o - 5s6s~ ^3\!S_1$  &  1.983   &  1.970  & -0.66\%  &  1.962(10)$^{\rm a}$\\
$5s5p~^3\!P_1^o - 5s6s~ ^3\!S_1$  &  3.463   &  3.439  & -0.70\%  &  3.425(17)\\[0.3pc]

$5s5p~^3\!P_0^o - 5s5d~ ^3\!D_1$  &  2.474   &  2.460  & -0.57\%  &  2.450(24)$^{\rm a}$\\
$5s5p~^3\!P_1^o - 5s5d~ ^3\!D_1$  &  2.065   &  2.017  & -2.38\%  &  2.009(20)\\
$5s5p~^3\!P_1^o - 5s5d~ ^3\!D_2$  &  3.720   &  3.688  & -0.87\%  &  3.673(37)\\[0.3pc]

$5s5p~^3\!P_0^o - 5p^2~ ^3\!P_1$  &  2.587   &  2.619  &  1.22\%  &  2.605(26)$^{\rm a}$\\
$5s5p~^3\!P_1^o - 5p^2~ ^3\!P_0$  &  2.619   &  2.671  &  1.95\%  &  2.657(27)\\
$5s5p~^3\!P_1^o - 5p^2~ ^3\!P_1$  &  2.317   &  2.374  &  2.40\%  &  2.362(24)\\
$5s5p~^3\!P_1^o - 5p^2~ ^3\!P_2$  &  2.837   &  2.880  &  1.49\%  &  2.865(29)
\end{tabular}
\end{ruledtabular}
\begin{flushleft}
$^{\rm a}${Reference~\cite{SafPorSaf13}}.
\end{flushleft}
\end{table}

 In~\cite{SafPorSaf13}, we used the CI+all-order method to evaluate the static and dynamic
polarizabilities of the $5s^2\,^1\!S_0$ and $5s5p\,^3\!P_0^o$ states of Sr.
We found  that the $E1$ matrix elements for the transitions that give
dominant contributions to the $^3\!P_0^o$ polarizability are sensitive
to the higher-order corrections to the wave functions and other
corrections to the matrix elements beyond the random phase
approximation (RPA). We included the higher-order corrections in an \textit{ab initio} way
using the CI+all-order approach and also calculated several other
corrections beyond RPA.
The resulting value for the dc Stark shift of the Sr $^1\!S_0 -\, ^3\!P_0^o$ clock transition, 247.5~a.u.,
was found to be in excellent agreement with the experimental result 247.374(7) a.u.~\cite{MidFalLis12}.

In order to predict the accurate values for the dynamic part of blackbody radiation shift in Sr clock, which is
one of the largest sources of Sr clock systematic uncertainty,
 we have combined
our theoretical calculations with the experimental measurements of
the Stark shift \cite{MidFalLis12} and magic wavelength
\cite{LudZelCam08} of the $5s^2\,^1\!S_0 - 5s5p\,^3\!P_0^o$ transition
to determine very accurate recommended values for several relevant electric-dipole matrix elements \cite{SafPorSaf13}. Specifically, we were able to obtain accurate recommended values for the following most important transitions contributing to the $^3\!P_0^o$ polarizability:
$5s5p\,\,^3\!P_0^o -\, 5s4d\,\,^3\!D_1$
$5s5p\,\,^3\!P_0^o -\, 5s6s\,\,^3\!S_1$,
$5s5p\,\,^3\!P_0^o -\, 5s5d\,\,^3\!D_1$, and
$5s5p\,\,^3\!P_0^o -\, 5p^2\,\,^3\!P_1$.

In this work, we use our previous results, supplemented with theoretical CI+all-order+RPA values of the reduced matrix element ratios, to obtain recommended values for 8 transitions that give dominant contributions to the polarizability of the $5s5p~^3\!P_1^o$ state.
The results are summarized in Table~\ref{tab1}.

We assume that the transitions from even-parity states to the $5s5p~^3\!P_1^o$ state are calculated in
the CI+all-order+RPA approach with the same accuracy as similar transitions from even-parity states to the $5s5p~^3\!P_0^o$ state,
for which the recommended values were presented in~\cite{SafPorSaf13}. Then, for example, the recommended value of the
$\langle 5s4d\,\,^3\!D_2 ||D|| 5s5p\,\,^3\!P_1^o \rangle$ matrix element is obtained here from the CI+all-order+RPA ratio
$$\frac{\langle 5s4d\,\,^3\!D_2 ||D|| 5s5p\,\,^3\!P_1^o \rangle}{\langle 5s4d\,\,^3\!D_1 ||D|| 5s5p\,\,^3\!P_0^o \rangle}$$ multiplied
by the recommended value of the $\langle 5s4d\,\,^3\!D_1 ||D|| 5s5p\,\,^3\!P_0^o \rangle$ reduced matrix element.
In a similar manner we find all other matrix elements listed in~\tref{tab1}.
We assign the uncertainties to the new recommended values  based on the uncertainties of the corresponding matrix element involving the
$5s5p~^3\!P_0^o$ state.

As a additional check,  we also use a simple ratio between relativistic and
nonrelativistic reduced matrix element of the electric-dipole  operator $D$ valid in the $LS$ coupling approximation.
Since the dipole and spin operators  commute, we obtain~\cite{Sob79},
\begin{eqnarray}
&&\langle \gamma J L S ||D|| \gamma' J' L' S'\rangle =
 \delta_{SS'} \sqrt{(2J+1)(2J'+1)}   \nonumber \\
& \times & (-1)^{S+L+J'+1} \, \left\{
\begin{array}{ccc}
L  & J  & S  \\
J' & L' & 1
\end{array}
\right\} \,
\langle \gamma L S ||D|| \gamma' L' S\rangle,
\label{MatEl}
\end{eqnarray}
where $S$ is the total spin momentum of the atomic state, $L$ and $J$ are the orbital and total angular momenta,
and $\gamma$ stands for all other quantum numbers.
\begin{table}
\caption{\label{table3} Contributions to the $5s5p~^3\!P_1^o$ static scalar polarizability of Sr in a.u.
The dominant contributions to the valence polarizabilities are listed separately. The theoretical and experimental
\cite{RalKraRea11} transition energies are given in columns $\Delta E_{\rm th}$  and $\Delta E_{\rm expt}$. The
remaining contributions to valence polarizability are given in row Other. The total of the core and
$\alpha^{vc}$ terms is listed together in row Core + Vc.
 The dominant contributions to $\alpha_0$, listed in columns
$\alpha_0[\mathrm{A}]$ and $\alpha_0[\mathrm{B}]$, are calculated with the CI + all-order + RPA
matrix elements and theoretical [A] and experimental [B] energies \cite{RalKraRea11}, respectively. The dominant
contributions to $\alpha_0$ listed in column $\alpha_0[\mathrm{C}]$ are calculated with experimental energies and our
recommended values of the matrix elements given in~\tref{tab1}.}
\begin{ruledtabular}
\begin{tabular}{lccrrrc}
 \multicolumn{1}{l}{Contribution} & \multicolumn{1}{c}{$\Delta E_{\rm th}$}
& \multicolumn{1}{c}{$\Delta E_{\rm expt}$} & \multicolumn{1}{c}{$\alpha_0[\mathrm{A}]$}
&\multicolumn{1}{c}{$\alpha_0[\mathrm{B}]$} & \multicolumn{1}{r}{$\alpha_0[\mathrm{C}]$} \\
\hline \\ [-0.3pc]
$5s5p ~^3\!P_1^o - 5s4d ~^3\!D_1$& 3589    &  3655  &    75.3 &  74.0 &   71.9 \\
$5s5p ~^3\!P_1^o - 5s4d ~^3\!D_2$& 3656    &  3715  &   221.5 & 218.0 &  212.0 \\[0.3pc]

$5s5p ~^3\!P_1^o - 5s6s ~^3\!S_1$& 14484   &  14535 &    39.8 &  39.7 &  39.4 \\[0.3pc]

$5s5p ~^3\!P_1^o - 5s5d ~^3\!D_1$& 20472   &  20503 &     9.7 &   9.7 &   9.6 \\
$5s5p ~^3\!P_1^o - 5s5d ~^3\!D_2$& 20488   &  20518 &    32.4 &  32.3 &  32.1 \\[0.3pc]

$5s5p ~^3\!P_1^o - 5p^2 ~^3\!P_0$& 20807   &  20689 &    16.7 &  16.8 &  16.6 \\
$5s5p ~^3\!P_1^o - 5p^2 ~^3\!P_1$& 21020   &  20896 &    13.1 &  13.2 &  13.0 \\
$5s5p ~^3\!P_1^o - 5p^2 ~^3\!P_2$& 21300   &  21171 &    19.0 &  19.1 &  18.9 \\[0.3pc]

$5s5p ~^3\!P_1^o - 5s7s ~^3\!S_1$& 22869   &  22920 &    1.8 &   1.8 &   1.8 \\[0.3pc]

 Other                           &         &        &   38.3 &  38.3 &  38.3 \\[0.3pc]

 Core +Vc                        &         &        &    5.55&   5.55&   5.55\\
 Total                           &         &        &  473.2 & 468.4 & 459.2 \\
 Recommended                     &         &        &        &       & 459.2(3.8)
\end{tabular}
\begin{flushleft}
\end{flushleft}
\end{ruledtabular}
\end{table}

The results produced by this formula for the transitions to $5s4d~^3D_J$ states differ from recommended values listed in
Table~\ref{tab1} by only 0.15\% and 0.2\%.
These differences are substantially smaller than the quoted uncertainties of 0.5\%.
The $5s5p~^3P_1^o-5s4d~^3D_J$ transitions give dominant contributions to the
$5s5p~^3P_1^o$ polarizability. The differences between the use of  Eq.~(\ref{MatEl}) and
the CI+all-order ratio for the other transitions range from 0.05\% to 5.6\%. This demonstrates that
$LS$ coupling works reasonably well for the $5s5p~^3\!P_J^o$, $5s4d~^3\!D_J$, and $5s5d~^3\!D_J$ terms.

We find that absolute values of all recommended matrix elements are slightly less than the \textit{ab initio}
CI+all-order results. The difference, as it was discussed in Ref.~\cite{SafPorSaf13}, can be attributed to the small corrections
beyond RPA, such as the core-Brueckner, two-particle, structural radiation, and normalization corrections.
\begin{table*}
\caption{\label{Polariz} The dominant contributions to the Sr $5s5p~^3\!P_1^o$ and
 $5s5p~^3\!P_0^o$ scalar polarizabilities in a.u. and in percent. \textit{Ab initio} (columns 2-5) and recommended (columns 6-9) values are given.
Nonrelativistic term notation is used when the sum of relativistic
contributions is given (see the main text for details). The results for the $5s5p~^3\!P_0^o$ state are taken
from Ref.~\cite{SafPorSaf13}. Final (recommended) results for the scalar $5s5p~^3\!P_1^o$ polarizability are given in the last column.}
\begin{ruledtabular}
\begin{tabular}{lrrccrrccc}
\multicolumn{1}{l}{} &\multicolumn{4}{c}{Theor. matrix elements and energies}
&\multicolumn{4}{l}{Recomm. matrix elements and exp. energies} &\multicolumn{1}{c}{Recomm.} \\
\multicolumn{1}{l}{State} & \multicolumn{1}{r}{$\alpha_0(^3\!P_1^o)$} & \multicolumn{1}{r}{$\alpha_0(^3\!P_0^o)$}
& \multicolumn{1}{c}{$\alpha_0(^3\!P_1^o)$} & \multicolumn{1}{c}{$\alpha_0(^3\!P_0^o)$}
& \multicolumn{1}{r}{$\alpha_0(^3\!P_1^o)$} & \multicolumn{1}{r}{$\alpha_0(^3\!P_0^o)$}
& \multicolumn{1}{c}{$\alpha_0(^3\!P_1^o)$} & \multicolumn{1}{c}{$\alpha_0(^3\!P_0^o)$}
& \multicolumn{1}{c}{$\alpha_0(^3\!P_1^o)$} \\
\hline \\ [-0.3pc]
$5s4d~^3\!D$   & 296.8 & 285.0 & 62.7\% & 62.2\% & 283.9 & 272.6 & 61.8\% & 61.3\% & 283.9(3.4) \\
$5s6s~^3\!S_1$ &  39.8 &  38.7 &  8.4\% &  8.4\% &  39.4 &  38.3 &  8.6\% &  8.6\% &  39.4(0.4) \\
$5s5d~^3\!D$   &  42.1 &  42.9 &  8.9\% &  9.4\% &  41.7 &  42.5 &  9.1\% &  9.6\% &  41.7(0.8) \\
$5p^2~^3\!P$   &  48.8 &  47.3 & 10.3\% & 10.3\% &  48.6	&  47.1 & 10.6\% & 10.6\% &  48.6(0.9) \\
$5p7s~^3\!S_1$ &  1.81 &  1.70 &  0.4\% &  0.4\% &  1.81 &  1.69 &  0.4\% &  0.4\% &  1.81(0.05) \\[0.2pc]
Other          &  38.3 &  36.9 &  8.1\% &  8.1\% &  38.2 &  36.9 &  8.3\% &  8.3\% &  38.2(1.1)  \\[0.2pc]
Core+Vc        &   5.6 &   5.6 &  1.2\% &  1.2\% &   5.6 &   5.6 &  1.2\% &  1.2\% &  5.55(0.06) \\[0.3pc]
Total          & 473.2 & 458.1 &100.0\% &100.0\% & 459.2 & 444.6 &100.0\% &100.0\% & 459.2(3.8)
\end{tabular}
\end{ruledtabular}
\end{table*}

Along with the recommended values, we also give \textit{ab initio}
results of the CI+MBPT and CI+all-order calculations that include RPA corrections to the effective operator.
The higher-order (HO) corrections may be estimated as the difference of the  CI+all-order+RPA and CI+MBPT+RPA calculations.
These contributions of the higher orders, listed in the ``HO'' column of Table~\ref{tab1}, provide a good estimate of the
uncertainty and are larger than the  more accurate final uncertainty estimate for most of the transitions.
Since the basis set is numerically complete and the configuration space is saturated for two electrons, the contribution to
the uncertainty budget coming from CI is negligible in comparison to the contributions arising from core-valence correlations.

\section{Polarizability of the $^3\!P_1^o$ state}
\label{Polariz1}

We calculated the static and dynamic polarizabilities of the Sr $5s5p~^3\!P_1^o$ state  using the high-precision CI+all-order method. The dynamic polarizability $\alpha(\omega)$
can be represented as a sum
\begin{equation}
\alpha(\omega) = \alpha^v(\omega) + \alpha^c(\omega) + \alpha^{vc}(\omega),
\end{equation}
where $\alpha^v(\omega)$ is the valence polarizability, $\alpha_c$ is
the ionic core polarizability, and a small term $\alpha^{vc}$ compensates for Pauli-principle forbidden excitations to
occupied valence shells and slightly modifies the ionic core polarizability.

The valence part of the polarizability is determined  by solving the inhomogeneous
equation in valence space, which is approximated as~\cite{KozPor99a}
\begin{equation}
(E_v - H_{\textrm{eff}})|\Psi(v,M^{\prime})\rangle = D_{\textrm{eff}} |\Psi_0(v,J,M)\rangle
\label{inho}
\end{equation}
for the state  $v$ with total angular momentum $J$ and magnetic quantum number $M$. The parts of the wave function $\Psi(v,M^{\prime})$ with angular momenta of $J^{\prime}=J,J \pm 1$ allow us to determine the scalar and tensor
polarizabilities of the state $|v,J,M\rangle$ \cite{KozPor99a}.
The effective dipole operator $D_{\textrm{eff}}$ includes RPA corrections.

 Small core terms $\alpha^c$ and $\alpha^{vc}$ are evaluated in the RPA. The latter
is calculated by adding $\alpha^{vc}$ contributions from the individual electrons, i.e.,
$\alpha^{vc}(5s5p)=\alpha^{vc}(5s)+\alpha^{vc}(5p)$. The uncertainties of these terms are
determined by comparing the Dirac-Fock and RPA values.

We use the sum-over-states formula for
the scalar part of the dynamic valence polarizability~\cite{MitSafCla10} to establish the dominant contributions to the final value
\begin{equation}
\alpha_0^v(\omega) = \frac{2}{3(2J+1)}\sum_n\frac{(E_n-E_v)|\langle v\| D\| n\rangle|^2}{(E_n-E_v)^2-\omega^2} .
\label{genpol}
\end{equation}
Here $J$ is the total angular momentum of the state $v$ and $E_n$ is the energy of the state $n$.
For the static polarizability, $\omega = 0$ in~\eref{genpol}. Determination of the dominant contributions is essential for
estimating the uncertainty of the final value.

We have carried out several calculations of the dominant contributions to the $5s5p~^3\!P_1^o$ static scalar polarizability
using different sets of the energies and $E1$ matrix elements. The results are presented in Table~\ref{table3}. The theoretical and experimental \cite{RalKraRea11} transition energies  are given in
columns $\Delta E_{\rm th}$ and $\Delta E_{\rm expt}$ in cm$^{-1}$. The dominant contributions to the polarizability listed in columns
$\alpha_0[\mathrm{A}]$ and $\alpha_0[\mathrm{B}]$ are calculated with CI + all-order + RPA  matrix elements and theoretical [A]
and experimental [B] energies~\cite{RalKraRea11}, respectively.  The dominant contributions to $\alpha_0$ listed in column
$\alpha_0[\mathrm{C}]$  are calculated with experimental energies and recommended  matrix elements. These results are taken as final.
The remaining valence contributions that are not listed separately are given in row labeled ``Other''.
The sum of the core and $\alpha^{vc}$ terms is listed in row labeled ``Core +Vc''.

A comparison of the main contributions to the $5s5p ~^3\!P_0^o$  and
$5s5p ~^3\!P_1^o$ scalar polarizabilities is given in~\tref{Polariz}. Two sets of calculations are presented.
In the first calculation, we use the {\it ab initio} values of the matrix elements and  energies, while in the second calculation the recommended matrix elements and the experimental energies are used.
To simplify the comparison, we sum the contributions from the transitions to
$^3\!D_J$ and $^3\!P_J$ states and use $^3\!D$ and $^3\!P$ terms labels for the totals.
For example, the contribution of the intermediate $5s4d~^3\!D$ state to $\alpha_0(^3\!P_1^o)$ means the sum of contributions
of the $5s4d~^3\!D_1$ and $5s4d~^3\!D_2$ states, while in the case of the $^3\!P_0^o$ state the notation
$5s4d~^3\!D$ means the contribution of the $5s4d~^3\!D_1$ state only.

The contribution of different terms to $\alpha_0(^3\!P_1^o)$ and $\alpha_0(^3\!P_0^o)$ is very similar.
Therefore,  we are able to assign the uncertainties to these contributions based on the uncertainties
of the matrix elements listed in~\tref{tab1} and on the uncertainties of the respective contributions to $\alpha_0(^3\!P_0^o)$
determined in~\cite{SafPorSaf13}. Our final recommended result for the $5s5p~^3\!P_1^o$ scalar polarizability
is 459.2(3.8) a.u..
\section{Magic wavelength}
\label{mw}
The magic wavelength $\lambda^*$ at which $\alpha_{^1\!S_0}(\lambda^*) = \alpha_{^3\!P^o_{1,|m_J|=1}}(\lambda^*)$
and the quadratic Stark shift on the $^1\!S_0-\,^3\!P_1^o$ transition vanishes, was experimentally determined by
Ido and Katori~\cite{IdoKat03} to be 914(1) nm. Note that $\alpha_{^3\!P^o_{1,|m_J|=1}}$ is the {\it total} polarizability,
i.e., is the sum of the scalar and tensor parts.
Using the magic frequency $\omega^*= 0.049851(5)\,\, {\rm a.u.}$, corresponding to the magic wavelength $\lambda^*$,
and the experimental value of the matrix element $|\langle 5s^2 \,^1\!S_0 ||D|| 5s5p \,^1\!P_1^o \rangle| = 5.248(2)$ a.u.~\cite{YasKisTak06},
$\alpha_{^1\!S_0}(\omega^*)$ was obtained in Ref.~\cite{PorLudBoy08} to be 261.2(3) a.u..

Solving inhomogeneous equation,~\eref{inho}, we found $\alpha_{^1\!S_0}(\omega^*)$ = 261.03 a.u. in excellent agreement with the result 261.2(3).
When this value was recalculated with the recommended matrix elements and the experimental energies we obtained 261.07 a.u..
Similar calculations  of the  $\alpha_{^1\!P^o_{1,|m_J|=1}}(\omega*)$ yield 264.3 a.u. and 261.0 a.u., respectively. Therefore, the use of
the recommended matrix elements and the experimental energies  yields the experimentally  determined magic wavelength to within its
stated uncertainty.
\section{$C_6$ coefficients}
\label{Sec:C6}
The expression for
the $C_{6}(^1\!S_0 +\, ^3\!P_1^o)$ coefficient is given by \cite{PorSafDer14}
\begin{equation}
C_{6}(\Omega _{p})=\sum_{J=0}^{2}A_{J}(\Omega )X_{J},
\label{C6}
\end{equation}
where the angular dependence $A_J(\Omega)$ is represented by
\begin{equation}\label{eq1}
A_{J}(\Omega)=\frac{1}{3}\sum_{\mu =-1}^{1}\left\{ w_{\mu }^{(1)}
\left(
\begin{array}{ccc}
1 & 1 & J \\
-\Omega & -\mu & \Omega +\mu%
\end{array}%
\right) \right\} ^{2}
\end{equation}
with the dipole weights $w_{\pm 1}^{(1)}=1$ and $w_0^{(1)}=2$ and $\Omega =0,1$.
The coefficients $A_J(\Omega)$ (and, consequently,
the $C_6$ coefficients) do not depend on gerade/ungerade symmetry.

The quantities $X_J$ for the $^1\!S_0 +\, ^3\!P_1^o$ dimer are given by
\begin{equation}
X_J = \frac{27}{2\pi} \int_0^\infty \alpha_1^A(i\omega) \,
\alpha^B_{1J}(i\omega) \, d\omega + \delta X_0 \, \delta_{J,0} .
\label{X_J}
\end{equation}
where $A \equiv \, ^1\!S_0$ and  $B \equiv\, ^3\!P_1^o$,
possible values of the total angular momentum $J$ are 0, 1, and 2,
and the other quantities are defined below.

The
$\alpha_1^A(i\omega)$ is the electric-dipole dynamic polarizability of the $^1\!S_0$ state
at the imaginary argument.
The quantity $\alpha^\Phi_{KJ}(i\omega)$ is a part of the
scalar electric-dipole ($K=1$) or electric-quadrupole ($K=2$) dynamic polarizability of
the state $\Phi$, in which the sum over the intermediate states $|n \rangle$ is restricted
to the states with fixed total angular momentum $J_n=J$:

\begin{eqnarray}
&&\alpha^\Phi_{KJ}(i\omega) \equiv \frac{2}{(2K+1)(2J_\Phi+1)}  \nonumber \\
&\times&\!\sum_{\gamma_n} \frac{(E_n - E_\Phi)
|\langle \gamma_n, J_n=J ||T^{(K)}|| \gamma_\Phi, J_\Phi \rangle|^2} {(E_n - E_\Phi)^2 + \omega^2} .
\label{alphaB}
\end{eqnarray}
Here $\gamma_n$ stands for all quantum numbers of the intermediate states except $J_n$.

The correction $\delta X_0$ to the $X_0$ term in~Eq.(\ref{X_J})
arises due to a downward $^3\!P_1^o \rightarrow \,^1\!S_0$ transition
and is given by the following expression:
\begin{eqnarray}
\delta X_0 &=& 2\,|\langle ^{3}\!P_{1}^{o}||D||^{1}\!S_{0}\rangle|^2
\sum_{n\neq \,^3\!P_1^o}
\frac{(E_n-E_{^1\!S_0})\,
|\langle n||D||^1\!S_0 \rangle|^2}{(E_n-E_{^1\!S_0})^2-\omega_0^2} \nonumber \\
&+& \frac{|\langle ^3\!P_1^o||D||^{1}\!S_{0}\rangle |^{4}}{2\omega_0} ,
\label{delX0}
\end{eqnarray}
where $\omega_0= E_{^3\!P_1^o}-E_{^1\!S_0}$.

A breakdown of the $C_6(\Omega)$ contributions for the Sr $(^1\!S_0 +\, ^3\!P_1^o)$ dimer is given in Table~\ref{tabA}.
\begin{table*}[tbph]
\caption{A breakdown of the contributions to the $C_6(\Omega)$ coefficient for the $(^1\!S_0 +\, ^3\!P_1^o)$ dimer.
The CI+MBPT+RPA values for $X_J$ are given in column labeled ``CI+MBPT''. The explanation for two other calculations
listed in ``CI+All'' and ``Recomm.'' columns is given in the text. The $\delta X_0$ term is
given separately in the second row; it is included in the $J=0$ contribution. }
\label{tabA}
\begin{ruledtabular}
\begin{tabular}{lccccccccc}
\multicolumn{1}{l}{} &\multicolumn{2}{c}{$A_J$} &\multicolumn{3}{c}{$X_J$}
&\multicolumn{2}{c}{$C_6$ [CI+All]} &\multicolumn{2}{c}{$C_6$ [Recomm.]} \\
\multicolumn{1}{l}{$J$} &\multicolumn{1}{c}{$\Omega=0$}& \multicolumn{1}{c}{$\Omega=1$}
&\multicolumn{1}{c}{CI+MBPT} &\multicolumn{1}{c}{CI+All}  &\multicolumn{1}{c}{Recomm.}
&\multicolumn{1}{c}{$\Omega=0$}& \multicolumn{1}{c}{$\Omega=1$}&\multicolumn{1}{c}{$\Omega=0$}& \multicolumn{1}{c}{$\Omega=1$}\\
\hline \\[-0.6pc]
$0$         & 4/9   & 1/9    & 1473  &  1494 & 1486  &  664 &  166 &  660     &  165 \\
$\delta X_0$& 4/9   & 1/9    & 22.8  &  23.7 &   23  &   11 &    3 &   10     &    3 \\
$1$         & 1/9   & 5/18   & 6406  &  6395 & 6320  &  711 & 1776 &  702     & 1756 \\
$2$         & 11/45 & 19/90  & 9981  & 10007 & 9853  & 2446 & 2113 & 2409     & 2080 \\
Sum         &       &        &       &       &       & 3821 & 4055 & 3771     & 4001 \\
Recommended &       &        &       &       &       &      &      & 3771(32) & 4001(33)
\end{tabular}
\begin{flushleft}
\end{flushleft}
\end{ruledtabular}
\end{table*}
Two calculations were carried out:
\begin{itemize}
\item In the first calculation (labeled ``CI+All'' in \tref{tabA}) the CI+all-order+RPA values of matrix elements and energies
were used for $\alpha_1(^3\!P_1^o)(i\omega)$. For $\alpha_1(^1\!S_0)(i\omega)$ we used the experimental $^1\!S_0 -\, ^1\!P_1^o$
electric dipole matrix element and experimental transition energy for all frequencies.

\item In the second calculation (labeled ``Recomm.'' in \tref{tabA})
the CI+all-order matrix elements and energies were replaced by the recommended matrix elements
and the experimental energies for {\it all} frequencies in the evaluation of  $\alpha_1(^3\!P_1^o)(i\omega)$.
\end{itemize}

We list in Table~\ref{tabA} the quantities $X_J$ and coefficients $A_J$  given by Eqs.~(\ref{eq1}) and (\ref{X_J})
for allowed $J=0, 1, 2$. The $\delta X_0$ term is given separately in the second row to illustrate the magnitude of this contribution.
It is very small, 0.3\% of the total for $\Omega=0$ and 0.07\% for $\Omega=1$.

The fractional uncertainty $\delta C_6$ for the $A+B$ dimer may be expressed via
fractional uncertainties in the scalar static dipole polarizabilities of the atomic states $A$ and $B$~\cite{PorDer03},
\begin{equation}
 \delta C_6  \approx
 \sqrt{\left( \delta \alpha_1^A(0) \right)^2 +
       \left( \delta \alpha_1^B(0)\right)^2} .
\label{AB}
\end{equation}

The polarizabilities and their absolute uncertainties are presented in~\tref{C_68}.
\begin{table}[tbph]
\caption{The $5s^2\,^1\!S_0$, $5s5p\,^3\!P^o_0$, and $5s5p\,^3\!P^o_1$ electric-dipole,
$\alpha_1$, static polarizabilities in the
CI+MBPT+RPA and CI+all-order+RPA approximations are given in columns labeled ``CI+MBPT''
and ``CI+All''. For the $^3\!P^o_1$ state
the scalar ($\alpha_{1s}$) and tensor ($\alpha_{1t}$) parts of the polarizabilities are presented.
$C_6(\Omega_{u/g})$ coefficients for the $A+B$ dimers  are listed in the bottom part.
The values of $C_6$, given in column ``CI+All'', were obtained with the CI+all-order+RPA values of $\alpha_{1s}(^3\!P_1^o)(i\omega)$
and CI+all-order+RPA values of $\alpha_1(^1\!S_0)(i\omega)$
(adjusted for the experimental $^1\!P_1^o -\,^1\!S_0$ matrix element and transition energy).}
\label{C_68}%
\begin{ruledtabular}
\begin{tabular}{lccccr}
\multicolumn{1}{c}{Level}& \multicolumn{1}{c}{Property}&
\multicolumn{1}{c}{CI+MBPT}& \multicolumn{1}{c}{CI+all}&\multicolumn{1}{c}{HO}&
\multicolumn{1}{r}{Recomm.}\\
\hline \\ [-0.6pc]
$5s^2\,^1\!S_0$         & $\alpha_1$       &  195.4  &  197.8  &  1.2\% & 197.14(20)$^{\rm a}$ \\
$5s5p\,^3\!P_0^o$       & $\alpha_1$       &  482.1  &  458.1  & -5.2\% & 444.51(20)$^{\rm a}$ \\
$5s5p\,^3\!P_1^o$       & $\alpha_{1s}$    &  499.1  &  473.2  & -5.2\% & 459.2(3.8) \\
                        & $\alpha_{1t}$    &   27.9  &   25.7  & -8.6\% &  26(2)  \\ [0.2pc]
$^1\!S_0 +\, ^3\!P_1^o$ & $C_6(0_{u\!/g})$ &  3806   &  3821   &  0.4\% & 3771(32)        \\
                        & $C_6(1_{u\!/g})$ &  4050   &  4055   &  0.1\% & 4001(33)
\end{tabular}
\end{ruledtabular}
\begin{flushleft}
$^{\rm a}${From Ref.~\cite{SafPorSaf13}.}
\end{flushleft}
\end{table}
The uncertainty of the electric-dipole static $^1\!S_0$ polarizability was
discussed in detail in Ref.~\cite{SafPorSaf13}; its recommended value is
$\alpha_0(^1\!S_0) = 197.14(20)$ a.u.. The uncertainty of the
scalar static $^3\!P^o_1$ polarizability was determined in this work to be 0.8\%.
The uncertainty of the tensor part of the static $^3\!P^o_1$ polarizability was determined
as the difference of the CI+all-order+RPA and CI+MBPT+RPA values.
Using the uncertainties of the scalar polarizabilities and~Eq.(\ref{AB}) we are able to determine the fractional uncertainty
of the $C_6(\Omega)(^1\!S_0 +\, ^3\!P_1^o)$ coefficients to be 0.83\%. The final recommended values
are presented in~\tref{tabA}.

\section{Electric quadrupole polarizabilities}
\label{C8}
In this section we discuss the calculation of  the static electric quadrupole polarizabilities for
the $5s^2\,^1\!S_0$ and $5s5p\,^3\!P_1^o$ states.
There are three contributions to $\alpha_2(^3\!P_1^o)$ coming from the intermediate states with $J=1,2,3$.

Using~\eref{alphaB}, we find for the valence part of the reduced dynamic scalar electric quadrupole polarizability $\alpha_2(i\omega)$
of the state $|\gamma_0, J_0\rangle \equiv |0\rangle$ with the energy $E_0$
\begin{equation}
\alpha^v_{2J}(i \omega) = \frac{2}{5(2J_0+1)} \sum_n \frac{(E_n - E_0)
|\langle \gamma_n J_n=J ||Q|| 0 \rangle|^2} {(E_n - E_0)^2 + \omega^2} .
\label{alpha2}
\end{equation}

To correctly include the core contributions for all projections $J$ we use the equation
\[
\alpha_{2 J}=\alpha^v_{2 J} + \frac{2J+1}{5(2J_0+1)}(\alpha^c_2+\alpha^{vc}_2),
\]
from ~\cite{PorSafDer14}, where we assume that the factor $(2J+1)/(5(2J_0+1))$ is the same for both
$\alpha^c_2$ and $\alpha_2^{vc}$. This is correct for
the $^1\!S_0$ state. For the $^3\!P_1^o$ state, the $\alpha_2^{vc}$ term is negligibly small.

The breakdown of the contributions to the $5s^2~^1\!S_0$ $E2$ and $5s5p~^3\!P_1^o$ $E2$ scalar
polarizabilities obtained using the CI+MBPT and CI+all-order methods is given in~\tref{tableE2}.
\begin{table*}[tbph]
\caption{\label{tableE2} Contributions to the $5s^2~^1\!S_0$ and  $5s5p~^3\!P_1^o$ quadrupole scalar  polarizabilities.
The experimental transition energies~\cite{RalKraRea11} are given in column $\Delta E_{\rm expt}$.
Theoretical transition energies, absolute values of electric quadrupole reduced matrix elements $Q$,
and the dominant contributions to $\alpha_2$ are given for the CI+MBPT+RPA and CI+all-order+RPA approximations
in columns labeled $\Delta E_{\rm th}$ (in cm$^{-1}$), $Q$ (in a.u.), and $\alpha_2$ (in a.u.).
The remaining contributions to valence polarizability are grouped together in row Other. The contributions from the core and
$vc$ terms are listed together in row Core + Vc.
The recommended values of the dominant contributions to $\alpha_2$, listed in column $\alpha_2$[Recom],
are calculated with the CI+all-order+RPA values of $Q$ and the experimental energies.}
\begin{ruledtabular}
\begin{tabular}{llcccrrrrr}
&&& \multicolumn{3}{c}{CI+MBPT+RPA} & \multicolumn{3}{c}{CI+all-order+RPA} &  \\
\multicolumn{1}{l}{State} &\multicolumn{1}{l}{Contribution} & \multicolumn{1}{c}{$\Delta E_{\rm expt}$}
& \multicolumn{1}{c}{$\Delta E_{\rm th}$} & \multicolumn{1}{c}{$Q$} & \multicolumn{1}{r}{$\alpha_2$}
& \multicolumn{1}{c}{$\Delta E_{\rm th}$} & \multicolumn{1}{c}{$Q$} &\multicolumn{1}{r}{$\alpha_2$}
& \multicolumn{1}{r}{$\alpha_2$[Recom]} \\
\hline \\ [-0.6pc]
$5s^2 ~^1\!S_0$   &$5s^2 ~^1\!S_0 - 5s4d ~^3\!D_2$    & 18394  & 18298  &  1.20 &     7 & 18219  &  1.18  &     7 &     7  \\
                  &$5s^2 ~^1\!S_0 - 5s4d ~^1\!D_2$    & 20441  & 20428  & 26.00 &  2905 & 20150  & 26.54  &  3026 &  3069 \\
                  &$5s^2 ~^1\!S_0 - 5s5d ~^1\!D_2$    & 34958  & 35092  & 17.39 &   757 & 34727  & 17.26  &   749 &   754  \\
                  &$5s^2 ~^1\!S_0 - 5s5d ~^3\!D_2$    & 35226  & 35387  &  0.52 &     1 & 35022  &  0.54  &     1 &     1  \\
                  & Other                             &        &        &       &   689 &        &        &   697 &   697 \\[0.3pc]

                  & Core +Vc                          &        &        &       &    17 &        &        &    17 &    17 \\
                  & Total                             &        &        &       &  4375 &        &        &  4496 &  4545 \\
                  & Recommended                       &        &        &       &       &        &        &       &  4545(120) \\[0.4pc]

$5s5p ~^3\!P_1^o$ &$5s5p ~^3\!P_1^o - 5s5p~^3\!P_2^o$ &   394.2&   404.3& 36.30 & 95397 &   403.2& 36.46  & 96487 & 98680 \\
                  &$5s5p ~^3\!P_1^o - 4d5p~^3\!F_2^o$ & 18763  & 18724  & 15.51 &   376 & 18909  & 16.17  &   404 &   408 \\
                  &$5s5p ~^3\!P_1^o - 4d5p~^3\!F_3^o$ & 19085  & 19095  & 25.46 &   994 & 19264  & 26.26  &  1048 &  1057 \\
                  &$5s5p ~^3\!P_1^o - 5s6p~^3\!P_1^o$ & 19364  & 19260  &  9.89 &   149 & 19332  & 10.07  &   154 &   153 \\
                  &$5s5p ~^3\!P_1^o - 5s6p~^3\!P_2^o$ & 19469  & 19370  & 16.03 &   388 & 19395  & 18.61  &   522 &   521 \\
                  & Other                             &        &        &       &  4596 &        &        &  4482 &  4482 \\[0.3pc]

                  & Core +Vc                          &        &        &       &    17 &        &        &    17 &    17 \\
                  & Total                             &        &        &       &101917 &        &        &103114 &105317 \\
                  & Recommended                       &        &        &       &       &        &        &       &1.053(12) $\times 10^5$
\end{tabular}
\begin{flushleft}
\end{flushleft}
\end{ruledtabular}
\end{table*}
The RPA corrections to the quadrupole operator were also included. The recommended values obtained by replacing the theoretical transition energies
by the experimental ones are given in the last column of the table. The uncertainties were determined as the differences of the CI+all-order+RPA and CI+MBPT+RPA results.

The main contribution to the ground state quadrupole  polarizability   comes from the $5s4d ~^1\!D_2$ and $5s5d ~^1\!D_2$
states, which give together 84\% of total.
The main contribution to the scalar part of the $^3\!P^o_1$ static quadrupole polarizability comes from
the $5s5p ~^3\!P^o_2$ state. This intermediate state gives 94\% of total. This is due to a very small energy interval
$\Delta E = E(^3\!P^o_2) - E(^3\!P^o_1)$ of only 394.2 cm$^{-1}$. We  note that we obtained very close results for $\Delta E$,
404 and 403 cm$^{-1}$, at the CI+MBPT and CI+all-order stages, respectively. At the same time these values are 2.5\% larger
than the experimental transition energy $\Delta E =$ 394.2 cm$^{-1}$. This difference is taken into account in the
recommended values of the quadrupole polarizabilities, where the experimental energies are used for the dominant transitions.

\section{$C_8$ coefficients}
\label{C8sub}
The $C_8$ dispersion coefficient for the $^1\!S_0 +\, ^1\!S_0$ dimer can be
found as the quadrature of the electric-dipole $\alpha^A_1(i \omega)$ and
the electric-quadrupole $\alpha^A_2(i \omega)$ dynamic polarizabilities of the $A \equiv\, ^1\!S_0$ state:
\begin{eqnarray}
C_8 &=& \frac{15}{\pi}\, \int_0^\infty\, \alpha_1^A(i \omega)\,
\alpha_2^A(i \omega)\, d\omega .
\label{C8_1S0}
\end{eqnarray}

The results of calculation of the $C_8$ coefficient in the CI+MBPT+RPA
and CI+all-order+RPA approximations are presented in Table~\ref{X_JaJb}.
\begin{table*}[tbp]
\caption{The $X_k^{J_a J_b}$ for different $J_a$, $J_b$, and $k$
and $C_8$ coefficients obtained in the CI+MBPT+RPA and CI+all-order+RPA approximations are
given in columns labeled ``CI+MBPT'' and ``CI+All''.
The recommended values, given in column labeled ``Recomm.'',
are calculated with CI+all-order+RPA values of $Q$ and the experimental transition energies to the
mainly contributing intermediate states listed in~\tref{tableE2}. The contribution of $\delta X^{11}_1$ is included in $X^{11}_1$
and the contribution of $\delta X^{20}_2$ is included in $X^{20}_2$.
The (rounded) recommended values are taken as final.
Higher-order contributions, defined as relative differences of the ``CI+All'' and ``CI+MBPT'' values,
are listed in column labeled ``HO'' in \%. Determination of uncertainties, given in parenthesis,
is discussed in the text.}
\label{X_JaJb}%
\begin{ruledtabular}
\begin{tabular}{lcrrrrc}
                      &                    & CI+MBPT & CI+All &    HO      & Recomm. & Final \\
\hline \\[-0.6pc]
$^1\!S_0+\,^1\!S_0$   & $C_8$              & 361454  & 370965 &   2.4\%    & 371455 & $3.7(1) \times 10^5$ \\[0.3pc]

$^1\!S_0+\,^3\!P_1^o$ & $\delta X^{11}_1$  & 126234  & 128515 &   1.8\%    & 128515 &\\
                      & $X^{11}_1$         & 186311  & 189088 &   1.5\%    & 189088 &\\
                      & $X^{12}_1$         & 704597  & 713986 &   1.3\%    & 711015 &\\
                      & $X^{13}_1$         & 425433  & 429976 &   1.1\%    & 429976 &\\[0.3pc]

                      & $\delta X^{20}_2$  &    863  &    890 &   3.0\%    &    920 &\\
                      & $X^{20}_2$         &  57407  &  59132 &   2.9\%    &  59202 &\\
                      & $X^{21}_2$         & 247138  & 249097 &   0.8\%    & 249812 &\\
                      & $X^{22}_2$         & 383908  & 387224 &   0.9\%    & 388487 &\\[0.3pc]

                      & $X^{11}_3$         &         &     90 &            &     90 &\\[0.3pc]
                      & $X^{22}_4$         &         &    154 &            &    154 &\\[0.3pc]

                      & $C_8(0_u)$         & 555172  & 561944 &   1.2\%    & 562406 & $5.624(73) \times 10^5$ \\
                      & $C_8(1_u)$         & 724902  & 733370 &   1.2\%    & 732694 & $7.327(86) \times 10^5$ \\[0.3pc]
                      & $C_8(0_g)$         & 554954  & 561727 &   1.2\%    & 562187 & $5.622(73) \times 10^5$ \\
                      & $C_8(1_g)$         & 724829  & 733297 &   1.2\%    & 732621 & $7.326(86) \times 10^5$ \\
\end{tabular}
\end{ruledtabular}
\end{table*}
The recommended value is obtained with the CI+all-order+RPA matrix elements of the $Q$ operator and the experimental transition energies
for the intermediate states listed in~\tref{tableE2}. The recommended value is taken as final.

In analogy to the $C_6$ coefficient the fractional uncertainty of $C_8(^1\!S_0 +\, ^1\!S_0)$ can be expressed via
fractional uncertainties of the electric dipole and quadrupole static polarizabilities of the $A\equiv \,^1\!S_0$ state as
\begin{equation}
 \delta C_8(^1\!S_0 +\, ^1\!S_0)  \approx
 \sqrt{\left( \delta \alpha_1^A(0) \right)^2 +
       \left( \delta \alpha_2^A(0)\right)^2} .
\label{delC8_A}
\end{equation}
Now taking into account that $\delta \alpha_1^A(0)$ is negligible in comparison to $\delta \alpha_2^A(0)$,
we arrive at $\delta C_8 \approx \delta \alpha_2^A(0) \approx 2.6\%$.

The $C_8(5s^2\,^{1}\!S_{0}+5s5p\,^{3}\!P_{1}^{o})$ coefficient can be written in a general form~\cite{PorSafDer14}:
\[
C_8(\Omega_p)= \sum_{l=1}^4 \sum_{J_\alpha J_\beta}
A_l^{J_\alpha J_\beta}(\Omega_p) X_l^{J_\alpha J_\beta} .
\]

The non-zero angular factors $A_l^{J_\alpha J_\beta}(\Omega_p)$  are listed in~\tref{A_JaJb}.
\begin{table}
\caption{The values of the $A_l^{J_{a}J_{b}}(\Omega_{p})$ coefficients. The parameter $p=0$ for
ungerade symmetry and $p=1$ for gerade symmetry.}
\label{A_JaJb}%
\begin{ruledtabular}
\begin{tabular}{ccc}
           & $\Omega_p=0$    & $\Omega_p=1$ \\
\hline
$A^{11}_1$ &      3/5        &  1/5   \\
$A^{12}_1$ &      1/15       &  7/15  \\
$A^{13}_1$ &     43/105      & 31/105  \\[0.2pc]
$A^{20}_2$ &      3/5        &  1/5    \\
$A^{21}_2$ &      1/5        &  2/5    \\
$A^{22}_2$ &      9/25       &  8/25   \\[0.2pc]
$A^{11}_3$ &  $(-1)^p$ 3/5   &  $(-1)^p$ 1/5   \\
$A^{22}_4$ &  $(-1)^p$ 9/25  &  $(-1)^p$ 3/25
\end{tabular}
\end{ruledtabular}
\end{table}
A derivation of the corresponding quantities $X_l^{J_\alpha J_\beta}$ was discussed in detail
in Ref.~\cite{PorSafDer14}, therefore, we give only the final formulas:
\begin{eqnarray}
X_1^{1\!J} &=& \frac{45}{2\pi}\, \int_0^\infty \alpha^A_1(i\omega)\,
\alpha^B_{2 J}(i\omega)\, d\omega + \delta X_{1}^{11}\, \delta_{J1}\,  \nonumber \\
\delta X_{1}^{11} &=& \frac{3}{2}\,|\langle ^3\!P_1^o||Q||^3\!P_1^o \rangle|^2 \, \alpha_1^A(0),
\label{X_1}
\end{eqnarray}
where $J=1, 2, 3$.
\begin{eqnarray}
X_{2}^{2J} &=& \frac{45}{2\pi }\,\int_{0}^{\infty }
\alpha_2^A(i\omega) \, \alpha^B_{1 J} (i\omega)\, d\omega +
\delta X_{2}^{20}\, \delta_{J0}\,  \nonumber \\
\delta X_{2}^{20} &=& 5 \,|\langle ^{3}\!P_{1}^{o}||D||^{1}\!S_{0}\rangle|^{2}
\alpha_2^A(\omega_0),
\label{X_2}
\end{eqnarray}%
where $J=0, 1, 2$  and $\omega _{0}\equiv E_{\,^{3}\!P_{1}^{o}}-E_{\,^{1}\!S_{0}}$.
\begin{eqnarray*}
X_{3}^{11}&=&\sum_{n,k}
\frac{ \langle^{1}\!S_{0}||D||n\rangle \langle n ||Q||^{3}\!P_{1}^{o}\rangle
\langle^{3}\!P_{1}^{o}||Q||k\rangle \langle k||D||^{1}\!S_{0}\rangle }
{E_{n}-E_{\,^{1}\!S_{0}}+E_{k}-E_{\,^{3}\!P_{1}^{o}}}, \nonumber \\
X_{4}^{22}&=&\sum_{n,k}
\frac{\langle ^{1}\!S_{0}||Q||n\rangle \langle n||D||^{3}\!P_{1}^{o}\rangle
\langle^{3}\!P_{1}^{o}||D||k \rangle \langle k||Q||^{1}\!S_{0}\rangle }
{E_{n}-E_{\,^{1}\!S_{0}}+E_{k}-E_{\,^{3}\!P_{1}^{o}}}.
\label{X_34}
\end{eqnarray*}
A complete calculation of the $X^{11}_3$ and $X^{22}_4$ terms is rather difficult due to double summations
over intermediate states $n$ and $k$. However, these expressions can be simplified if we note that
the main contributions to the static electric dipole and quadrupole $^1\!S_0$ polarizabilities come from
a few low-lying intermediate states. Thus, we can leave in the sums over index $k$ in $X^{11}_3$ and $X^{22}_4$ 
only a few first terms arriving at the following approximate expressions:
\begin{widetext}
\begin{eqnarray}
X_{3}^{11} &\approx&\langle ^{3}\!P_{1}^{o}||Q||5s5p\,^{3}\!P_{1}^{o}\rangle
\langle 5s5p\,^{3}\!P_{1}^{o}||D||^{1}\!S_{0}\rangle \sum_{n} \frac{\langle
^{1}\!S_{0}||D||n\,\rangle \langle n\,||Q||^{3}\!P_{1}^{o}\rangle} {%
E_{n}-E_{^{1}\!S_{0}}}  \nonumber \\
&&+\langle ^{3}\!P_{1}^{o}||Q||5s5p\,^{1}\!P_{1}^{o}\rangle \langle
5s5p\,^{1}\!P_{1}^{o}||D||^{1}\!S_{0}\rangle \sum_{n}\frac{\,\langle
^{1}\!S_{0}||D||n\,\rangle \langle n\,||Q||^{3}\!P_{1}^{o}\rangle } {%
E_{n}-E_{^{1}\!S_{0}}+E_{(5s5p\,^{1}\!P_{1}^{o})}-E_{\,^{3}\!P_{1}^{o}}}
\nonumber \\
&&+\langle ^{3}\!P_{1}^{o}||Q||5s6p\,^{3}\!P_{1}^{o}\rangle \langle
5s6p\,^{3}\!P_{1}^{o}||D||^{1}\!S_{0}\rangle \sum_{n}\frac{\,\langle
^{1}\!S_{0}||D||n\,\rangle \langle n\,||Q||^{3}\!P_{1}^{o}\rangle } {%
E_{n}-E_{^{1}\!S_{0}}+E_{(5s6p\,^{3}\!P_{1}^{o})}-E_{^{3}\!P_{1}^{o}}}
\nonumber \\
&&+\langle ^{3}\!P_{1}^{o}||Q||5s6p\,^{1}\!P_{1}^{o}\rangle \langle
5s6p\,^{1}\!P_{1}^{o}||D||^{1}\!S_{0}\rangle \sum_{n}\frac{\langle
^{1}\!S_{0}||D||n\,\rangle \langle n\,||Q||^{3}\!P_{1}^{o}\rangle } {%
E_{n}-E_{^{1}\!S_{0}}+E_{(5s6p\,^{1}\!P_{1}^{o})}-E_{\,^{3}\!P_{1}^{o}}}.
\label{X311}
\end{eqnarray}%
\begin{eqnarray}
X_{4}^{22} &\approx&\,\langle ^{1}\!S_{0}||Q||5s4d\,^{3}\!D_{2}\rangle
\langle 5s4d\, ^{3}\!D_{2}||D||^{3}\!P_{1}^{o}\rangle \sum_{k,J_k=2}\frac{%
\langle ^{3}\!P_{1}^{o}||D||k\,\rangle \langle k\,||Q||^{1}\!S_{0}\rangle } {%
E_{k}-E_{\,^{3}\!P_{1}^{o}}+E_{(5s4d\,^{3}\!D_{2})}-E_{\,^{1}\!S_{0}}}
\nonumber \\
&&+\langle ^{1}\!S_{0}||Q||5s4d\,^{1}\!D_{2}\rangle \langle 5s4d\,
^{1}\!D_{2}||D||^{3}\!P_{1}^{o}\rangle \sum_{k,J_k=2}\frac{\langle
^{3}\!P_{1}^{o}||D||k\,\rangle \langle k\,||Q||^{1}\!S_{0}\rangle }{%
E_{k}-E_{\,^{3}\!P_{1}^{o}}+E_{(5s4d\,^{1}\!D_{2})}-E_{^{1}\!S_{0}}}
\nonumber \\
&&+\langle ^{1}\!S_{0}||Q||5s5d\,^{1}\!D_{2}\rangle \langle
5s5d~^{1}\!D_{2}||D||^{3}\!P_{1}^{o}\rangle \sum_{k,J_k=2}\frac{\langle
^{3}\!P_{1}^{o}||D||k\,\rangle \langle k\,||Q||^{1}\!S_{0}\rangle } {%
E_{k}-E_{^{3}\!P_{1}^{o}}+E_{(5s5d\,^{1}\!D_{2})}-E_{^{1}\!S_{0}}}  \nonumber
\\
&&+\langle ^{1}\!S_{0}||Q||5s5d\,^{3}\!D_{2}\rangle \langle
5s5d\,^{3}\!D_{2}||D||^{3}\!P_{1}^{o}\rangle \sum_{k,J_k=2}\frac{\langle
^{3}\!P_{1}^{o}||D||k\,\rangle \langle k\,||Q||^{1}\!S_{0}\rangle } {%
E_{k}-E_{^{3}\!P_{1}^{o}}+E_{(5s5d\,^{3}\!D_{2})}-E_{^{1}\!S_{0}}}.
\label{X422}
\end{eqnarray}
\end{widetext}

The $X_l^{J_{a}J_{b}}$ values and $C_8(\Omega_{g/u})$ coefficients for the
$5s^2\,^{1}\!S_{0}+5s5p\,^{3}\!P_{1}^{o}$ dimer are given in Table~\ref{X_JaJb}.
The contributions of the $X^{11}_3$ and $X^{22}_4$ are very small, which is expected since
 these terms contain intercombination transition matrix elements for both $D$ and
$Q$ operators. Such matrix elements are equal to zero in nonrelativistic approximation. Relativistic corrections are small for
Sr and, correspondingly, $X^{11}_3$ and $X^{22}_4$ are four orders of magnitude smaller than the main
contributions coming from $X^{1J}_1$ and $X^{2J'}_2$ terms.

The recommended values are calculated with the CI+all-order+RPA values of $Q$ and the experimental transition energies for the
main intermediate states listed in~\tref{tableE2}.
The (rounded) recommended values are taken as final.
Higher-order contributions were defined as relative differences of the CI+all-order+RPA and CI+MBPT+RPA values.

To estimate the uncertainties of the $C_8$ coefficients we neglect small quantities $X^{11}_3$ and $X^{22}_4$.
Then, designating
\begin{eqnarray}
 C_1 &\equiv& \sum_{J=1}^3 A_1^{1J} X_1^{1J}, \nonumber \\
 C_2 &\equiv& \sum_{J=0}^2 A_2^{2J'} X_2^{2J'} ,
\end{eqnarray}
we can express the absolute uncertainty of $C_8^{AB}$ via absolute uncertainties of $C_1$ and $C_2$ as
\begin{eqnarray}
\Delta C^{AB}_8 \approx \sqrt{ \Delta C_1^2+  \Delta C_2^2 } .
\label{C_AB}
\end{eqnarray}

The fractional uncertainties in $C_1$ and $C_2$ can be expressed via
corresponding fractional uncertainties in the scalar static polarizabilities
\begin{eqnarray}
\delta C_1  &\approx&  \sqrt{\left( \delta \alpha_1^A(0) \right)^2 + \left( \delta \alpha_2^B(0)\right)^2}, \nonumber \\
\delta C_2  &\approx&  \sqrt{\left( \delta \alpha_2^A(0) \right)^2 + \left( \delta \alpha_1^B(0)\right)^2} .
\label{delC8}
\end{eqnarray}

We note that $X_1^{11}$ includes the additional term $\delta X_1^{11}$ (see~\eref{X_1}). Nevertheless the equation above
for $\delta C_1$ is valid if we assume that the uncertainty of $|\langle ^3\!P_1^o||Q||^3\!P_1^o \rangle|^2$ is approximately
the same as the uncertainty of the scalar part of $\alpha_2(^3\!P_1^o)$. This assumption is based on that the $5s5p\,\,^3\!P_2^o$ state
contributes $\sim$ 94\% to the scalar $\alpha_2(^3\!P_1^o)$, i.e., the uncertainty of $\alpha_2(^3\!P_1^o)$ is
mostly determined by the uncertainty of the matrix element $|\langle ^3\!P_1^o||Q||^3\!P_2^o \rangle|$.
The latter is assumed to be the same as the uncertainty of the matrix element
$|\langle ^3\!P_1^o||Q||^3\!P_1^o \rangle|= 20.9$ a.u..

The term $\delta X_2^{20}$ contributing
to $X_2^{20}$ gives only 0.25\% of total $C_2$ and, respectively, its contribution to the
uncertainty budget is negligible.

Using these formulas and knowing the fractional uncertainties of the polarizabilities we assign the uncertainties to
the final values of the $C_8$ coefficients presented in~\tref{X_JaJb}. It is worth noting that if we estimate the uncertainties
of the $C_8$ coefficients as the difference of the CI+all-order+RPA and CI+MBPT+RPA values, we obtain very close results.

\section{Summary}
In~\tref{all} we summarize the results obtained for the van der Waals coefficients in this work.
We also include  the $C_6$ long-range interaction coefficients for the $^1\!S_0-\,^1\!S_0$, $^1\!S_0-\,^3\!P_0^o$, and
$^3\!P_0^o-\,^3\!P_0^o$ dimers provided in Ref.~\cite{ZhaBisBro14}. Comparing our results with other data, where available,
we see very good agreement for all quantities listed in the table.
\begin{table}[tbp]
\caption{
The $C_6$ and $C_8$ coefficients for the Sr $^1\!S_0 +\, ^1\!S_0$, $^1\!S_0 +\, ^3\!P_0^o$, $^3\!P_0^o +\,^3\!P_0^o$, and
$^1\!S_0 +\, ^3\!P_1^o$ dimers (in a.u.). The uncertainties are given in parenthesis.}
\label{all}%
\begin{ruledtabular}
\begin{tabular}{lccc}
                         &   Property       &     This work             & Other results \\
\hline \\ [-0.5pc]
$^1\!S_0 +\, ^1\!S_0$    &  $C_6$           &                           & 3107(4)$^{\rm c}$ \\
                         &                  &                           & 3103(7)$^{\rm b}$ \\
                         &                  &                           & 2890$^{\rm d}$ \\[0.2pc]
                         &  $C_8$           &  3.7(1)$\times 10^{5}$    & 3.792(8)$\times 10^{5\,{\rm b}}$ \\
                         &                  &                           & 3.854$\times 10^{5\,{\rm d}}$ \\[0.2pc]
$^1\!S_0 +\, ^3\!P_0^o$  &  $C_6$           &                           & 5360(200)$^{\rm c}$ \\
                         &                  &                           & 5260(500)$^{\rm e}$ \\[0.2pc]
$^3\!P_0^o +\,^3\!P_0^o$ &  $C_6$           &                           & 3880(80)$^{\rm c}$ \\[0.2pc]
$^1\!S_0 +\, ^3\!P_1^o$  & $C_6(0_{u\!/g})$ &   3771(32)                & 3868(50)$^{\rm f}$ \\
                         & $C_6(1_{u\!/g})$ &   4001(33)                & 4085(50)$^{\rm f}$ \\[0.2pc]
                         & $C_8(0_u)$       & $5.624(73) \times 10^5$   & \\
                         & $C_8(1_u)$       & $7.327(86) \times 10^5$   & \\[0.2pc]
                         & $C_8(0_g)$       & $5.622(73) \times 10^5$   & \\
                         & $C_8(1_g)$       & $7.326(86) \times 10^5$   &
\end{tabular}
\end{ruledtabular}
\begin{flushleft}
$^{\rm a}${Reference~\cite{SafPorSaf13}.} \\
$^{\rm b}${Reference~\cite{PorDer06Z}.} \\
$^{\rm c}${Reference~\cite{ZhaBisBro14}.} \\
$^{\rm d}${Reference~\cite{MitBro03}.} \\
$^{\rm e}${Reference~\cite{SanChrGre04}.} \\
$^{\rm f}${Reference~\cite{BorMorCiu14}.} \\
\end{flushleft}
\end{table}

To conclude, we evaluated $E1$ transition amplitudes from the $5s5p~^3\!P_1^o$ state to the low-lying even-parity
states and the electric dipole and quadrupole static polarizabilities of the $5s5p~^3\!P_1^o$ state of atomic Sr.
We also calculated the $C_6$ and $C_8$ coefficients for the Sr $^1\!S_0 +\, ^1\!S_0$ and $^1\!S_0 +\, ^3\!P_1^o$ dimers.
Our recommended values of the long range dispersion coefficients $C_6(0_u)=3771(32)$ a.u. and
$C_6(1_u)= 4001(33)$ a.u. are in a good agreement with the experimental results $C_6(0_u) = 3868(50)$ a.u.
and $C_6(1_u) = 4085(50)$ a.u. obtained in Ref.~~\cite{BorMorCiu14}.  We confirm the experimental value for
the $^1\!S_0-\,^3\!P_1^o$ magic wavelength.
We have analyzed the accuracy of calculations and assigned the uncertainties to all presented quantities.

\section{Acknowledgement*}
We thank P. Julienne, M. Borkowski, and  R. Ciury{\l}o for helpful discussions. This research
was performed under the sponsorship of the US Department
of Commerce, National Institute of Standards and Technology,
and was supported by the National Science Foundation under
Physics Frontiers Center Grant No. PHY-0822671 and by the
Office of Naval Research. The work of S.G.P. was supported
in part by US NSF Grant No. PHY-1212442.

\end{document}